\RequirePackage{fix-cm}
\documentclass[twocolumn]{svjour3}          
\smartqed  
\usepackage{graphicx}
\usepackage{siunitx}
\usepackage{setspace}
\usepackage{hyphenat}

\begin{document}

\title{ The Fano Signature in the Optical Response of a Waveguide-excited Compound Plasmonic Nanoantenna} 


\author{Hardik Vyas \and Ravi Hegde  
}

\institute{   Department of Electrical Engineering,
              IIT Gandhinagar, Gujarat, India. 382355\\
              \email{hegder@iitgn.ac.in}}           

\maketitle

\begin{abstract}

While long\hyp{}range propagating plasmons have been extensively investigated for implementing on\hyp{}chip optical sensing platforms, waveguide\hyp{}excited localized surface plasmon resonance (LSPR) based sensing systems have not yet received much attention. Waveguide excitation and readout as an alternative to free-space light based single\hyp{}particle spectroscopy are particularly attractive for high\hyp{}throughput sensing and on-chip drug discovery platforms. Here we present a numerical investigation of the optical response of a waveguide-excited plasmonic dolmen-shaped compound nanoantenna that exhibits a Fano signature in its spectral response. Although only evanescently coupled to the waveguide, the compound nanoantenna is seen to induce a high-contrast extinction in the transmission spectrum of the waveguide. The compound plasmonic nanoantenna configuration presented here is of interest in hybrid photonic\hyp{}plasmonic sensing approaches with high\hyp{}throughput capabilities. 
\end{abstract}
\keywords{ Plasmonic Nanoantenna \and Compound Plasmonic Nanoantenna \and Dielectric Waveguide \and Fano Resonance \and Refractive index Sensing}

\section{Introduction}
\label{intro}
Conventional optical sensing techniques use electromagnetic energy coupled long range surface propagating plasmons~\cite{pberini2009,pberini2016} (thin metallic films) as a means of sensing refractive changes in the medium. Sensing can also be achieved by using plasmonic nanostructures. Unlike in long range surface propagating plasmons, they can concentrate electromagnetic energy to sub-wavelength scales~\cite{Maier2007} and the sensing capability can be tuned by controlling the nanopaticle, size, shape, material. Plasmonic nanostructures can be used to sense changes in the tiny volume adjoining the surface. The above property can be exploited for various applications such as chemical and biological sensing~\cite{Anker2008}, Surface Enhanced Raman Spectroscopy (SERS)~\cite{Moskovic2006,Jackson2004}, Surface Enhanced Infrared Absorption
Spectroscopy (SEIRA)~\cite{Kundu2008} and surface enhanced nonlinear harmonic
generation~\cite{Scalora2010}. However, it is difficult to characterize these structures individually as they possess sub-wavelength size (this is due to the diffraction limit). For simplifying measurements, two-dimensional arrays of such nanoantennae, or nanostructures suspended in a liquid medium, are simultaneously excited and their collective response is measured and the averaged response is then interpreted as the response of an individual nanostructure~\cite{Rodriguez-Fortuno2011,Lorente-Crespo2013}. 
Additionally, independent excitation and measurement of multiple nanostructures in parallel, a capability that would improve the overall throughput, is 
difficult~\cite{Kundu2008},~\cite{Scalora2010} in free-space excitation schemes. 

A hybrid plasmonic-photonic approach through which isolated nanoantennae can be excited via high-index dielectric waveguides with transverse dimensions of the order of half of the wavelength would enable the excitation and measurement of multiple nanostructures in parallel, and, in real time, including the possibility of using different wavelengths, amplitudes and phases for independently driving each nanostructure~\cite{Rodriguez-Fortuno2016}. Besides this, such structures can be implemented using silicon photonics technology; fabrication using standard semiconductor fabrication techniques (such as the Complimentary Metal Oxide Semiconductor CMOS process) enables high-volume and low-cost production~\cite{Lipson2005,Thomson2016}. By integrating electronic circuitry, additional functionality can also be achieved. 

Plasmonic nanostructures placed on top of high refractive index dielectric waveguides have been experimentally studied in the visible and near infrared regions for LSPR based sensing applications~\cite{BernalArango2012,Alepuz-Benaches2012,Fevrier2012,Chamanzar2013,Peyskens2015,Peyskens2016}.
Dipolar excitation of the nanostructure takes place by the fundamental Transverse Electric (TE) mode of the guided wave, having its field oriented transverse to the direction of wave propagation on the waveguide plane. Since the nanostructures are placed in the evanescent field region of the waveguide the interaction efficiency per nanostructure  is smaller compared to free-space excitation ($<$20\%  interaction efficiency typically~\cite{Peyskens2015}); as a result, small extinction ratios are observed in the transmission spectra. Besides this, the dipolar resonance is broad due to the lossy nature of the metals and radiation damping; this is undesirable for sensing applications. The amount of dip in the transmission spectrum (or the contrast) can be increased by using a chain of nanoresonators on the waveguide~\cite{BernalArango2012,Fevrier2012,Peyskens2016} or by embedding nanostructures inside the waveguide \cite{Castro-Lopez2015,Espinosa-Soria2016a}. Embedding of the nanoantenna inside the waveguide requires introducing cuts in the waveguide and disrupts the separation between the wet and dry layers. Using multiple nanoantennae defeats the primary purpose of conducting single particle spectroscopy. Small fabrication related differences will lead to further broadening of the spectral response in the case of multiple nanoantennae. 

Fano resonances are sharp, narrow, asymmetric profile resonances~\cite{Verellen2009,Lukyanchuk2010} suitable for applications such as biosensing~\cite{Wu2011} and optical switches~\cite{Chang2012}. They are obtained as a result of the interference between bright and dark modes of plasmonic excitations such as the electric dipolar LSPR behaving as the bright mode and higher order LSPRs such as magnetic dipole~\cite{Shafiei2013} or electric quadrupole~\cite{Yang2011,Liuu2009} behaving as the dark mode. The phenomenon is analogous to Electromagnetically Induced Transparency (EIT) observed in quantum optics: a high-transmission narrow peak is obtained within a broad spectral region with high absorption~\cite{Harris2008}. In recent work, Fano resonances have been obtained using compound plasmonic nanoresonators coupled to plasmonic waveguides~\cite{Lu2012,Chen2014,Binfeng2016,Zhang2016}. Hybrid plasmonic\hyp{}photonic structures having compound plasmonic nanoantennae placed on top of high index dielectric waveguides have also been reported to obtain Fano resonance~\cite{Ortuno2017a}, but the obtained transmission modulation does not exhibit sufficient contrast.  The objective of our work is to numerically investigate the Fano resonance arising in a plasmonic dolmen nanoantenna (a compound nanoantenna that has shown promise in a free-space excitation configuration) when it is excited via an underlying optical waveguide. We show that the compound plasmonic nanoantenna leads to a higher field enhancement and increases the sensitivity of the system to small localized refractive index changes. The linewidth of the resonances is reduced in comparison with simple plasmonic nanoantennae. The transmission contrast for the peaks is also improved in the spectral response as compared to a previous report that also observed the Fano signature in a waveguide-excited configuration~\cite{Ortuno2017a}. The paper is organized as follows: following this introduction, the optical responses of waveguides loaded with one and two antennas is first considered and then used to 
understand the response of the dolmen structure in section~\label{sec:geometery}; a parametric study of effect of geometrical variations in the hybrid structure and the results of a simple surface refractive index sensing scheme are considered in section~\label{sec:results} before concluding the paper. 

\begin{figure}[htbp]
    \centering
    \includegraphics[width=1\linewidth]{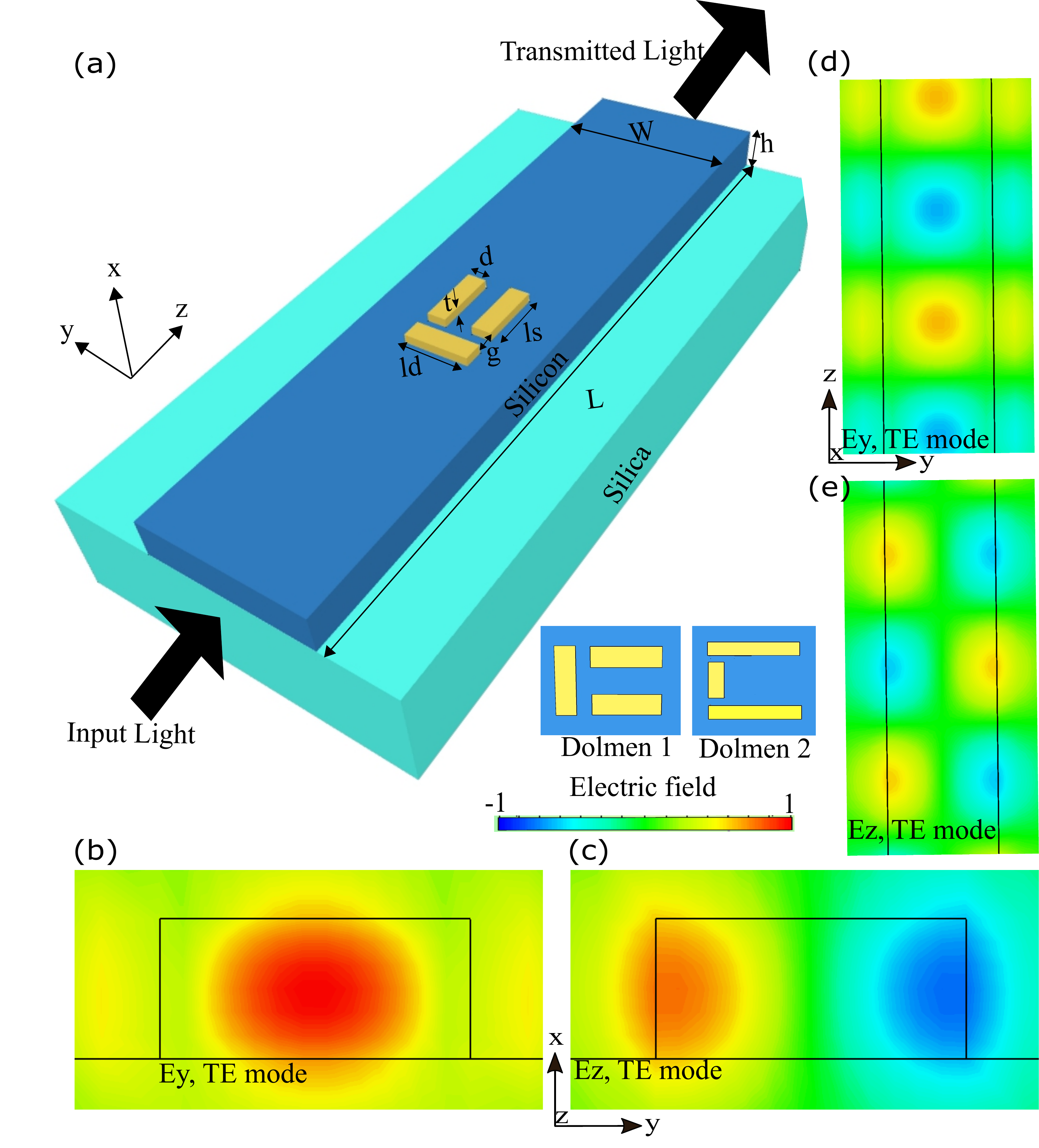}
    \caption{Schematic diagram of the plasmonic dolmen nanoantenna placed on top of a SOI waveguide. Two kinds of dolmen arrangements are possible as shown in the inlet. The normalized electric field distributions for the TE-mode at an excitation wavelength of \SI{1550}{\nano\meter} for the bare waveguide is shown for the $x-y$ cut in (b) and (c) and for the $y-z$ cut in (d) and (e) respectively. The waveguide is \SI{10}{\micro\meter} long, \SI{400}{\nano\meter} wide and \SI{180}{\nano\meter} thick throughout. Waveguide Core and substrate materials are Silicon and Silica respectively with air as background material.}
    \label{fig:schematic}
\end{figure}

\section{Geometry and Optical Response} \label{sec:geometry}

Figure~\ref{fig:schematic} (a) shows a Silicon On Insulator waveguide loaded with a single gold dolmen. In the initial study, we consider that the entire geometry is surrounded by a silica cladding. An air or water cladding are also of interest, but the influence of the cladding is to shift the center of the resonance. 
The dolmen nanoantenna consists of a single dipole bar and two quadrupole bars; the quadrupole bars can be arranged in two different ways as seen in the inlet of the figure. The simulation results are obtained using the Finite Integration Technique solver of the commercial solver CST Microwave studio. Open boundary conditions are imposed along all the axes. Gold is modeled using Johnson-Christy model~\cite{Johnson1972}. Refractive index values of 1.33, 1.45 and 3.45 are considered for modeling water, silica and silicon respectively. Firstly, the details of the bare silicon waveguide are considered.  
Figure~\ref{fig:schematic} (b-e) shows the cross sectional view of the modal field distribution for the bare waveguide. In particular, we note that there are both a transverse and a longitudinal component of the electric field.  In particular, it is seen that the $z$-component of the electric field achieves its maximum value near the periphery as opposed to the $y$-component. 

\begin{figure}[htbp]
    \centering
    \includegraphics[width=1\linewidth]{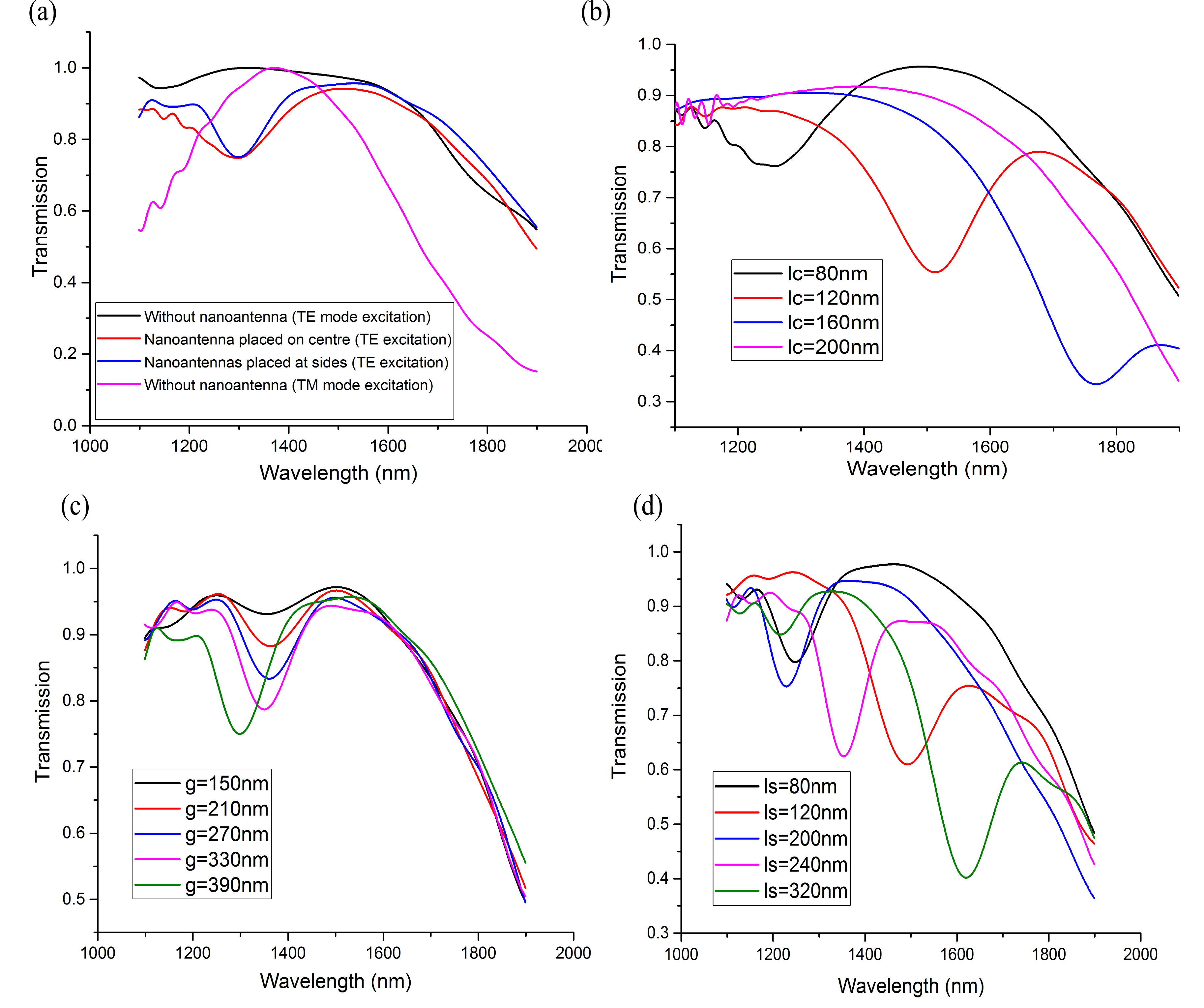}
    \caption{Numerically simulated transmission curves for the SOI waveguides loaded with one and two nanobar antenna. (a) shows the comparison of transmission spectra for the cases of nanoantenna loaded and bare silicon waveguides for TE mode excitation. Bare TM mode curve is shown for reference. The nanoantenna is bar shaped with the length of \SI{85}{\nano\meter}. (b) shows the influence of the dipole length for a waveguide loaded with a single nanoantenna in its center with TE mode excitation. (c) and (d) show the transmission curves when the waveguide is loaded with two symmetrically placed nanoantennae parallel to its edges. In (c), the separation between the two bars is varied and in (d), the lengths of the bars is varied. The width and heights of the bars are kept at \SI{30}{\nano\meter} throughout. The waveguide and background are modeled in a similar manner as in figure~\ref{fig:schematic}.}
    \label{fig:dipole}
    \end{figure}

When the bare waveguide is loaded with a bar-shaped nanoantenna, the evanescent fields at the interface of the high-index waveguide and the cladding excite the longitudinal dipolar plasmonic mode. Depending on the orientation of the long axis of the nanobar with respect to the propagation direction, the transverse and the longitudinal electric fields of the waveguide mode can be involved in the excitation.  In figure~\ref{fig:dipole}, the effect of dipolar resonance on transmission characteristics of the waveguide is studied. A transmission dip is observed in the presence of nanostrips as shown in figure~\ref{fig:dipole}(a). This is due to dipolar excitation of nanostrips by electric field components of the guided wave. The nanostrip with its length oriented along y-direction is excited by the transverse field component ($E_y$) of the guided wave. Similar dip is observed due to dipolar excitation of the symmetrically placed nanostrips by the longitudinal field component ($E_z$). Increasing the length of the nanostrips causes a red shift in the transmission spectrum as shown in figure~\ref{fig:dipole}(b),(c). However, in figure~\ref{fig:dipole}(b), no dip is observed on further increase in the nanostrip length. This is because the dipole extends into the waveguide edge where transverse component of the field is negligible as can be seen in the $E_y$ (TE mode) plot in figure~\ref{fig:schematic}(b). Besides this, for a given length of the nanostrips in both the cases, we observe an equal amount of transmission dip. Thus a single nanostrip placed at the waveguide centre gives a similar response as compared to the pair of nanostrips placed along the edges. In figure~\ref{fig:dipole}(d), reducing the spacing between the nanostrips reduces the amount of dip. This is because the nanostrips placed at the edges are brought closer to each other towards the waveguide centre, where the field strength of longitudinal component is negligible as can be seen in figure~\ref{fig:schematic} (c). 

\begin{figure}[htbp]
    \centering
    \includegraphics[width=1\linewidth]{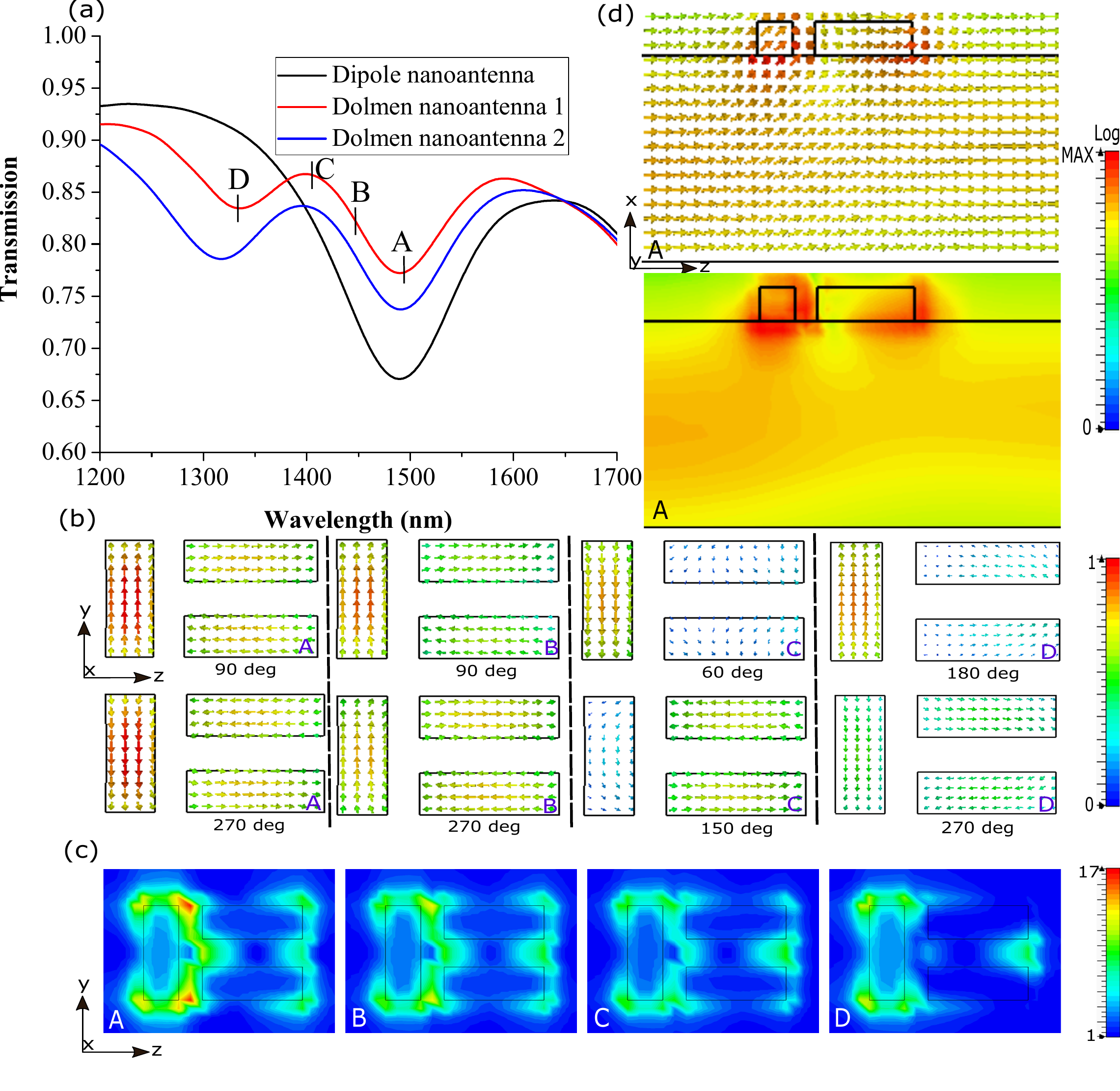}
    \caption{Fano resonance for the SOI waveguide loaded with plasmonic dolmen nanoantenna excited by TE-like mode. (a) shows Transmission curves for the waveguide loaded with dipole and the proposed dolmen nanoantennae. Wavelengths of interest are marked by points A,B,C,D. The dipole is \SI{103}{\nano\meter} long, \SI{50}{\nano\meter} wide and \SI{40}{\nano\meter} thick. For dolmen 1, each of the beams are \SI{85}{\nano\meter} long, \SI{30}{\nano\meter} wide and thick. The gap between the perpendicular and parallel beams is kept \SI{20}{\nano\meter}. For dolmen 2, the parallel beams are \SI{95}{\nano\meter} long, the perpendicular beam is \SI{75}{\nano\meter} long. Each beam is \SI{30}{\nano\meter} wide and thick. The gap between the perpendicular and parallel beams is kept \SI{20}{\nano\meter} for both the dolmen structures. (b)-(d) shows the near field results for the dolmen nanoantenna 1. (b) shows current density distributions on the surface of the dolmen at the points. For each wavelength, the distributions are shown at couple of time instants within one wavelength cycle. (c) shows electric field enhancements at the points. (d) shows cross-sectional power flow in logarithmic scale at wavelength corresponding to point C. Each of the nanoantennae are modeled using gold material (Johnson's model). The waveguide and background are modeled in a similar manner as in figure~\ref{fig:schematic}. }
    \label{fig:dolmen}
    \end{figure}
     
Next, we consider the modulation of the transmission spectrum of a waveguide loaded with the dolmen antenna in figure~\ref{fig:dolmen}. 
As seen in figure~\ref{fig:dolmen} (a), the response of the dolmen is different from that of the cases considered earlier. A Fano resonance signature is observed as a result of interference between dipole and quadrapole resonances. The Fano signature shows up as a high-transmission window within a broad low transmission peak. The transparency window is asymmetrical in shape in comparison with the typically symmetrical Lorentzian shape of the dipolar resonance. 
It is desired to relate the interference between dipole and quadrapole resonance in the near field with the far field transmission response.

In the near field plots of the current density and local electric field enhancement are shown in the figure~\ref{fig:dolmen}(b) and (c) respectively  at 4 
different wavelengths marked as A, B, C, D in figure~\ref{fig:dolmen}(a). For the current density plots, two time snapshots have been chosen each occuring at different points in a single wave period. At point A, the dipole and the quadrupole resonances are seen to peak at the same time instant. A minima is observed in the corresponding far field transmission response. At point C, we observe that the resonances occur nearly out of phase with each other; it is found that maximum current is observed in the quadrapole (represented by parallel beams) when the dipole (represented by perpendicular beam) shows minimum current and vice versa. It corresponds to a local maximum in the transmission response in the far field resulting from the destructive interference between the dipolar and the quadrupole resonances. At a point B, somewhere in between A and C, the phase difference is seen to be less than that of C and a reduced transmission in comparison to B is seen in the far field transmission spectrum. At Point D, only the presence of dipolar resonance can be seen. There is no significant quadrupole excitation as can be seen from the direction of current vectors in the parallel beams. It is also noted that similiar spectral features are seen for both the dolmen 1 and dolmen 2 variations with the overall transmission reduced in the case of dolmen 2, with the contrast being nearly identical. 
    
Figure~\ref{fig:dolmen}(c) represents the electric field enhancement observed around a dolmen nanoantenna (dolmen 1) at wavelengths corresponding to points A,B,C,D. It is found that the field enhancement is maximum around the corners of the dolmen structure. At point A, enhancement factor of 17 is observed near the inner edges of the parallel beam structure facing the perpendicular beam. The same region shows enhancement factor of 9 at wavelength corresponding to point B. For the remaining cases, maximum enhancement factor of 6 is observed. In the near field plots from points A to C, we observe a reduction in electric field enhancement. At point D, enhancement is observed near the dipole beam whereas negligible enhancement is observed near the quadrapole beams.  For the dolmen 2 variation, we note that at point A, the enhancement factor of 21 was observed. For the case of a single dipole antenna, the peak field enhancement factor was observed to be 14. Figure~\ref{fig:dolmen}(d) represents the cross-sectional power flow along the structure at the wavelength represented by point B. Considering the region near the core of the waveguide, reduction in power is observed as the guided wave propagates through the nanoantenna.

\section{Results and Discussions} \label{sec:results1}


\subsection{Influence of geometrical parameters}
\begin{figure}[ht]
    \centering
    \includegraphics[width=1\linewidth]{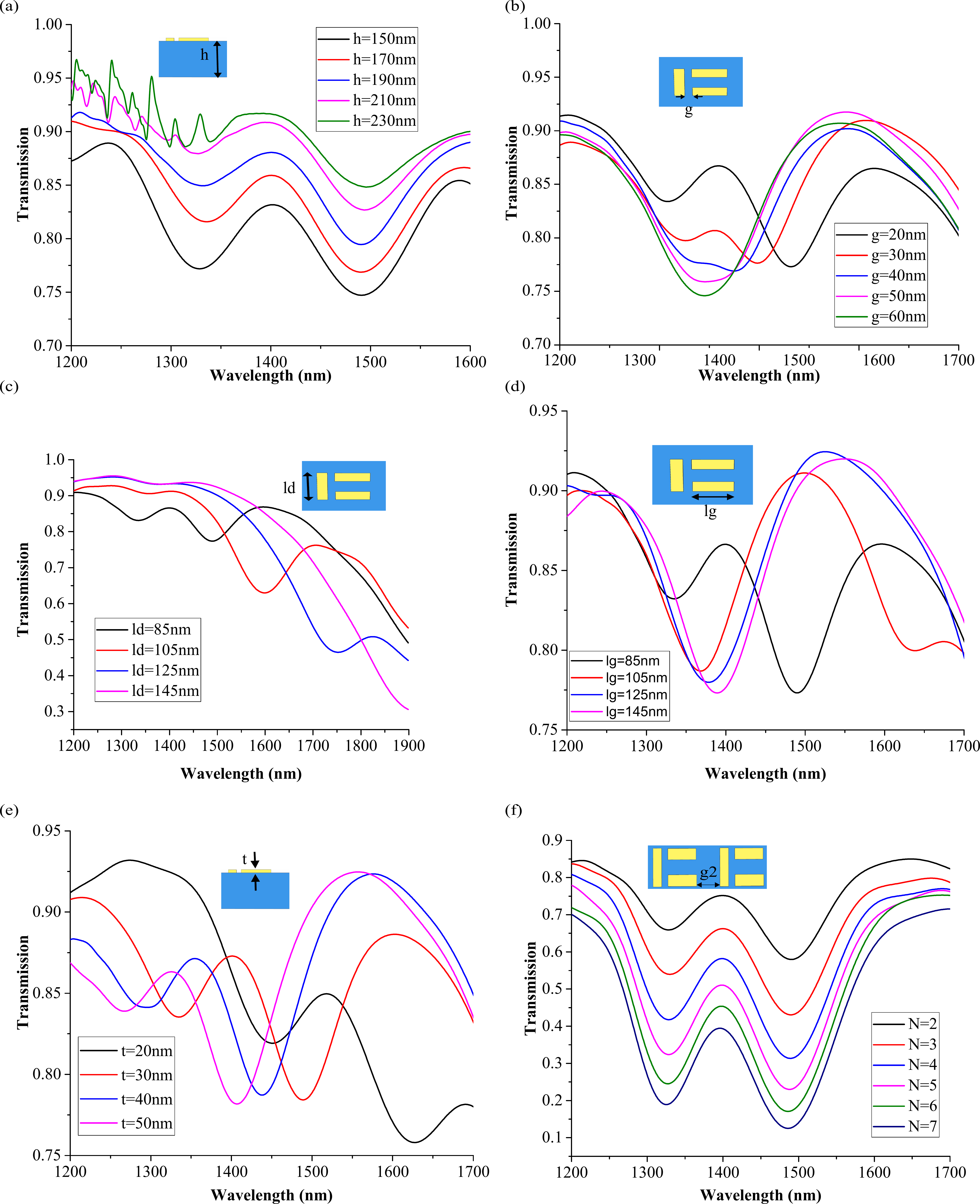}
    \caption{Effect of geometrical parameters of the waveguide and nanoantenna on transmission spectrum. (a) shows effect of changing waveguide thickness. For the dolmen nanoantenna, length of perpendicular beam and parallel beams is varied as shown in (c),(d). (b) shows variation of gap between the two beams, thickness of the dolmen structure is varied in (f). (e) shows effect of changing refractive index of the surrounding medium. (f) shows the effect of increasing the number of dolmen nanoantennae, the gap beteween the parallel and perpendicular beams is \SI{20}{\nano\meter} . The remaining dimensions of the waveguide, nanoantenna are same as that in figure~\ref{fig:dipole}. The effect of silica substrate has not been taken into account.}
    \label{fig:parametric}
    \end{figure}
    
    In this subsection, we discuss the means by which the spectral response of a dolmen-loaded waveguide can be tuned.  
Figures~\ref{fig:parametric}(a)-(f) represent geometrical parametric studies for a single plasmonic dolmen nanoantenna (dolmen 1) excited by a dielectric waveguide.
For the waveguide itself, the only parameter that can influence the spectrum shape and contrast is the height of the waveguide. In figure~\ref{fig:schematic}(a)  decreasing the height of the waveguide increases the interaction between the guided wave and the nanoantenna. As the height decreases, the overall transmission is also seen to drop. An improvement in the contrast between the peaks and the valleys in the spectrum is also seen as the height decreases. The near-field interaction between the dipole bar and the quadrupole bars is a direct mechanism to increase the contrast. In figure~\ref{fig:parametric}(b), we observe a change from the asymmetric Fano resonance to symmetric dipole resonance when the gap between them increases. Smaller and smaller gaps will also increase the local field enhancement. Increase in length of perpendicular beam decreases the dip due to dipolar resonance, increases the Fano dip and also causes red shift as shown in figure~\ref{fig:parametric}(c). The change in dip is in accordance with the results in figure~\ref{fig:dipole}(b),(d). In figure~\ref{fig:parametric}(d), increase in length of parallel beams causes a red shift in the Fano dip. However, there is no change in dip corresponding to the dipolar resonance. Increasing the thickness of the nanoantenna causes a blue shift in the transmission response as shown in figure~\ref{fig:parametric}(e). It is thus seen that by tuning the geometrical parameters, we can adjust the spectrum with considerable control. 

\begin{figure}[htbp]
    \centering
    \includegraphics[width=1\linewidth]{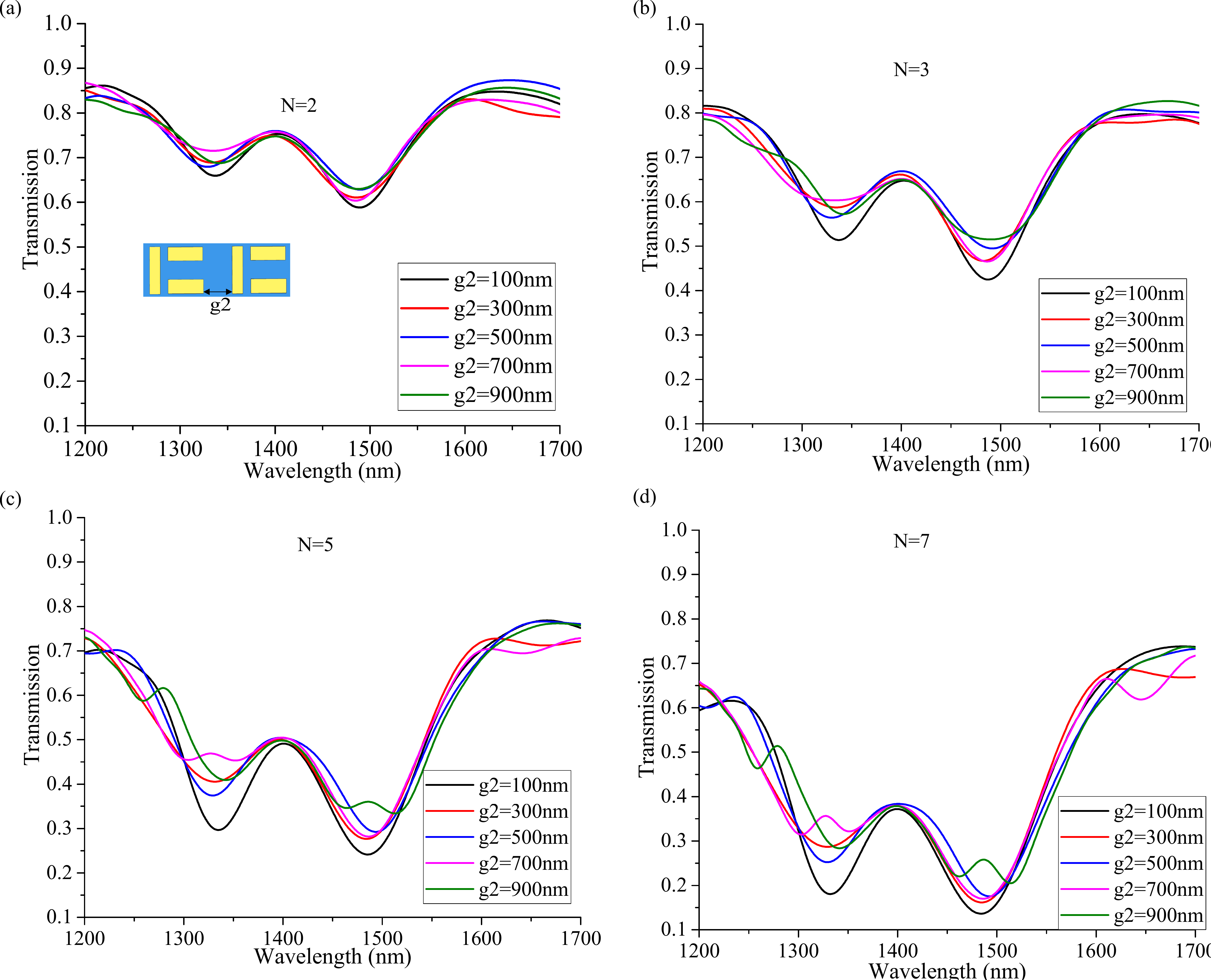}
    \caption{Effect of increasing the gap between adjacent nanoantennae (g2) on transmission spectrum  for a waveguide loaded with chain of nanoantennae. The waveguide and dolmen nanoantenna have same material and dimensions as that in figure~\ref{fig:dipole}.}
    \label{fig:nanoantennae gap}
    \end{figure}

Previous reports have studied contrast enhancement techniques by increasing the number of antennae along the waveguide. Even with the dolmen structure, contrast enhancement can be achieved by this technique. Two parameters can be tuned: the gap between the individual dolmens and the number of dolmen. 
In figure~\ref{fig:parametric} (f), it is observed that increasing the number of nanoantennae increases the Fano dip in transmission response and also improves the contrast. In figures~\ref{fig:nanoantennae gap}(a)-(d), a small decrease in the Fano dip is observed on increasing the gap between adjacent nanoantenna.
    
\subsection{Refractive index sensing}
Refractive index sensing for very small analyte volumes is one of the important applications of LSPRs~\cite{Rodriguez-Fortuno2011,Lukyanchuk2010,pberini2008}. Fano resonant compound plasmonic nanonantennae have been reported to exhibit higher sensitivities to local changes in refractive index in comparison to single nanoantennae. We consider the case of surface refractive index change sensing. It has been carried out for the following cases of nanoantennae loaded SOI waveguides: dipole, dolmen 1 and dolmen 2. A very thin layer of analyte is assumed to be deposited on the nanoantennae. The analyte layer is modeled such that it extends from the top and side nanoantennae surface to 20nm distance away from it. We are interested in determining both the optimal geometrical parameters of the dolmen itself as well as the optimal excitation wavelength at which a large change in the waveguide transmission can be observed for refractive index changes in the analyte layer. Sensing can be performed at or near the dipole or the quadrupole resonance wavelengths.  

\begin{figure}[htbp]
\centerline{\includegraphics[width=1\linewidth]{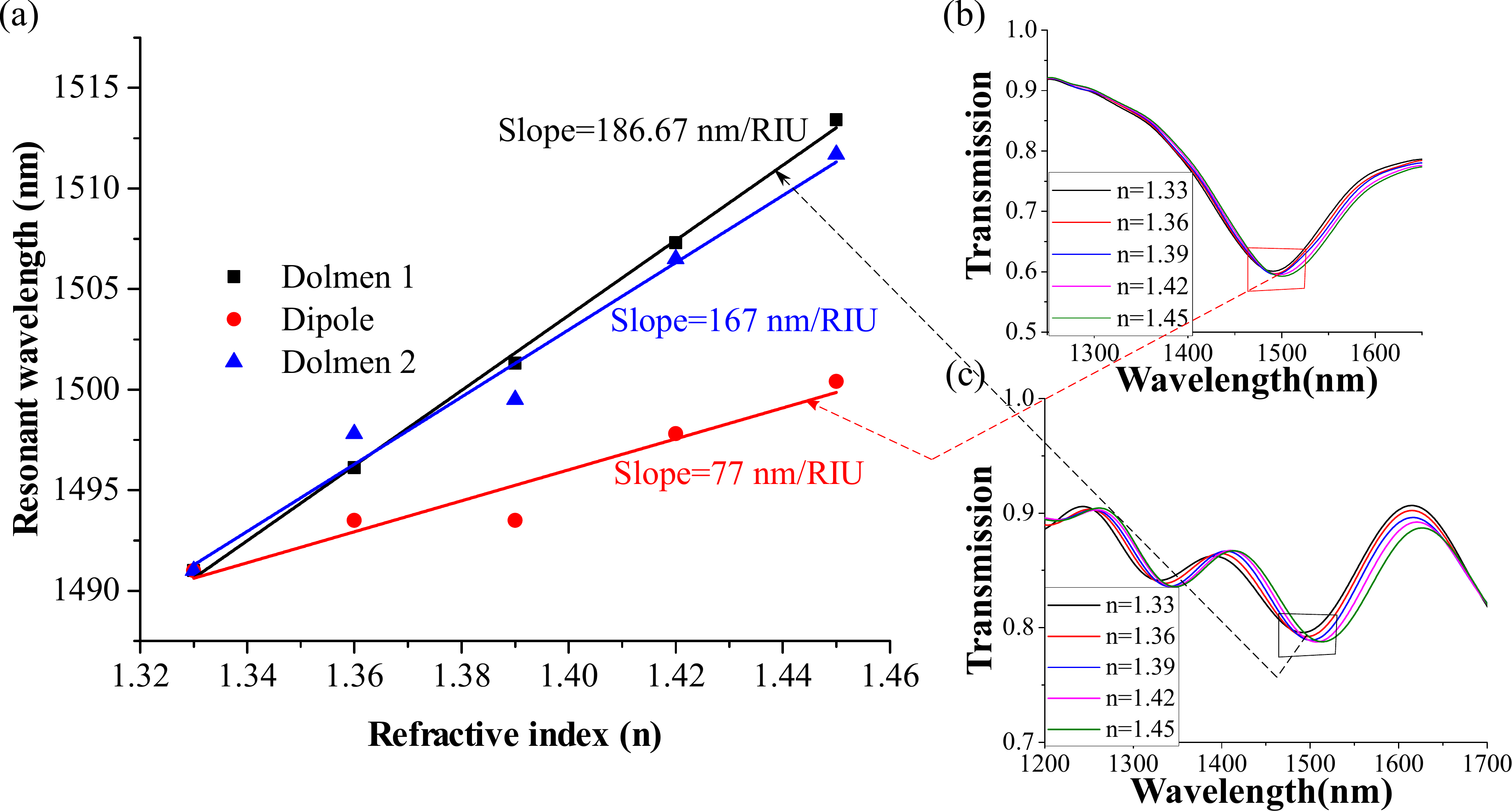}}
\caption{Comparision of surface refractive index sensing done using plasmonic dipole and dolmen nanoantenna placed on top of SOI waveguide. (a) shows the effect of changing refractive index on wavelength shift is shown for either type of sensing for the dipole and dolmen nanoantenna. (b), (c) shows the effect of changing refractive indices on transmission spectra corresponding to the dipole and dolmen nanoantenna 1 based waveguide structure respectively. The waveguide and the nanoantennae have same material and dimensions as that in figure~\ref{fig:dipole},~\ref{fig:dolmen}. }
\label{fig:RI sensing}
\end{figure}

Figures~\ref{fig:RI sensing} (b)-(c) gives a clear indication  of the effect of changing refractive index on transmission spectrum. The results shown in the figures correspond to dipole and dolmen 1 nanoantennae respectively. A red shift in the transmission spectra is observed with increase in refractive index in all the cases. In figure~\ref{fig:RI sensing} (a), the wavelengths corresponding to spectral peaks is seen for all the three cases of nanoantennae. The slopes in the figures give the values of sensitivity. Sensitivity is the ratio of change in resonance wavelength to the change in refractive index of the medium to be sensed.  Sensitivity is \SI{77}{\nano\meter}/RIU  for the dipole, whereas the dolmen 1 and 2 give the sensitivity of \SI{167}{\nano\meter}/RIU and \SI{186.67}{\nano\meter}/RIU respectively. 
Thus, for surface refractive index sensing, an improvement in sensitivity is observed when loading the waveguide with dolmen nanoantenna instead of dipole bar.

\section{Conclusion} 
In conclusion, we have studied the modulation of the transmission spectra of a dielectric waveguide when it loaded with a compound plasmonic nanonantenna. 
The Fano resonance occuring in the dolmen nanoantenna is shown to improve the sensitivity by a factor of nearly 2 to changes in the refractive index of a thin surrounding layer.  While the obtained contrasts in our case are nearly a factor of 3 in comparison with an earlier report, it still needs improvement. Further studies are investigating mechanisms to improve the interaction efficiency.

\section*{Acknowledgment}
The authors acknowledge support from the Department of Science and
Technology, Govt. of India through the Extramural grant SB/S3/EECE/0200/2015. 

\bibliographystyle{spmpsci}      
\bibliography{all1}
\end{document}